\documentstyle[preprint,aps]{revtex}

\begin{document}
\draft

\title{The role of the $S_{31}$ $\pi N$ partial wave in the three-nucleon 
force}

\author{T-Y. Saito}
\address{Research Center for Nuclear Physics, Osaka University, 
Ibaraki, Osaka 567, Japan}

\author{I. R. Afnan}
\address{Department of Physics, The Flinders University of South 
Australia, Bedford Park, SA 5042, Australia}

\date{\today}

\maketitle

\begin{abstract}

We re-examine the contribution of the $\pi N$ $S_{31}$ channel 
to the three-nucleon force for an energy dependent separable potential.
Despite the lack of cancellation between the $S_{31}$ and
the $P_{11}$ and $P_{33}$ channels, the total contribution of the 
three-nucleon force  to the triton binding  is still small.

\end{abstract}

\newpage

The very small contribution of the three-body force to the triton binding
energy  as recently reported by Saito and Afnan (SA) \cite{SA94,SA95} was
attributed to:  (i) The energy dependence of the $\pi N$ amplitude in the
subthreshold region.  (ii) The cancellation between the repulsive $S_{31}$
$\pi N$ partial wave and the attractive $P_{11}$ and $P_{33}$ partial waves. 
(iii) The very soft $\pi NN$ form factor for the pion emission and absorption
vertices. In this brief report  we re-examine in more detail the role of the
$S_{31}$ $\pi N$ amplitude in the cancellation between the contribution from
the different partial waves within the framework of a separable potential. 

In the previous studies\cite{SA94,SA95} the $\pi N$ potentials\cite{AWT76,MA85}
used were constructed to fit the scattering data with potential application to
$\pi d$ scattering. Here, we will concentrate on the subthreshold behavior
of the amplitude for a rank one separable potential with the aim of examining
how this behavior effects the contribution of the $S_{31}$ partial wave to
the three-nucleon force.

To get a better understanding of the role of the subthreshold behavior of
the amplitude for a rank one separable potential, we recall that the commonly
used potentials $v(k,k')$ are of the form
\begin{equation}
v(k,k') = g(k) \, \lambda \, g(k'),                                \label{eq1}
\end{equation}
where $g(k)$ is a form factor and $\lambda$ is the strength of the potential. 
The off-energy-shell t-matrix $T(k,k';E)$, is determined by solving the
Lippmann-Schwinger  equation
\begin{equation}
T(k,k';E) = v(k,k') + 
\int_0^\infty \, dq \,q^2 \, v(k,q)G(q;E)T(q,k';E)\ ,              \label{eq2}
\end{equation}
where the $\pi N$ propagator $G(q;E)$, is given as 
\begin{equation}
G(q;E) = {1 \over E - ( q^2/2m_N + m_N + \sqrt{q^2 + m_\pi^2}) }\ .\label{eq3}
\end{equation}
By solving Eq.(\ref{eq2}) for the above separable potential, the resultant
off-shell t-matrix takes the simple form
\begin{equation}
T(k,k';E) = g(k)\tau(E)g(k')\ ,                                    \label{eq4}
\end{equation}
where the energy dependence of the off-shell t-matrix is given by
\begin{equation}
\tau(E) = {1 \over 1/\lambda 
- \langle g | G(E) | g \rangle}\ ,                                \label{eq5}
\end{equation}
with
\begin{equation}
\langle g|G(E)|g \rangle = \int_0^\infty \, dq \, q^2 
g(q) \, G(q;E) \, g(q)\ .                                         \label{eq6}
\end{equation}
This t-matrix, $T(k,k';E)$, has a specific energy dependence in  the
subthreshold  region, which is most simply illustrated by considering the
derivative of $\tau(E)$ with respect to $E$, {\it i.e.},
\begin{equation}
{d \tau(E) \over d E} = - \left[\tau(E)\right]^2 \, 
\langle g | G^2(E) | g \rangle\ .                                 \label{eq7}
\end{equation}
Since $\tau(E)$ is real in the subthreshold region, $d \tau(E) /d E$ is
negative.  As a result, the t-matrix has negative slope as a function of $E$
for both  attractive and repulsive partial waves. 
For the $S_{31}$ $\pi N$ potential used previously\cite{SA94,SA95}, 
the value of
$\tau(E)$ increases, and approaches $\lambda = +1$  as $E\rightarrow\infty$.
To reduce the possible cancellation between the repulsive $S_{31}$ and the
attractive $\pi N$ partial waves, we propose to introduce a parameterization of
the $S_{31}$ potential that is energy  dependent, and in this way change the
energy dependence of the subthreshold amplitude within the framework of a
rank one separable potential.

This energy dependence in the potential is introduced by replacing the strength
of the potential $\lambda$, by  $c/(E - M)$ where $c$ and $M$ are parameters of
the potential. This allows $\tau(E)$ to deviate from the condition dictated by
Eq.~(\ref{eq7}). For  the form factor
$g(k)$, we use 
\begin{equation}
g(k) = {1 \over \sqrt{\omega_k}}\left({1\over k^2 + \alpha_1^2} + 
{\beta k^2 \over (k^2 + \alpha_2)^2}\right)\ ,                   \label{eq8}
\end{equation}
where $\omega_k = \sqrt{k^2 + m_\pi^2}$. The factor of 
$\frac{1}{\sqrt{\omega_k}}$ is introduced to maintain consistency with the
relativistic treatment of the pion.

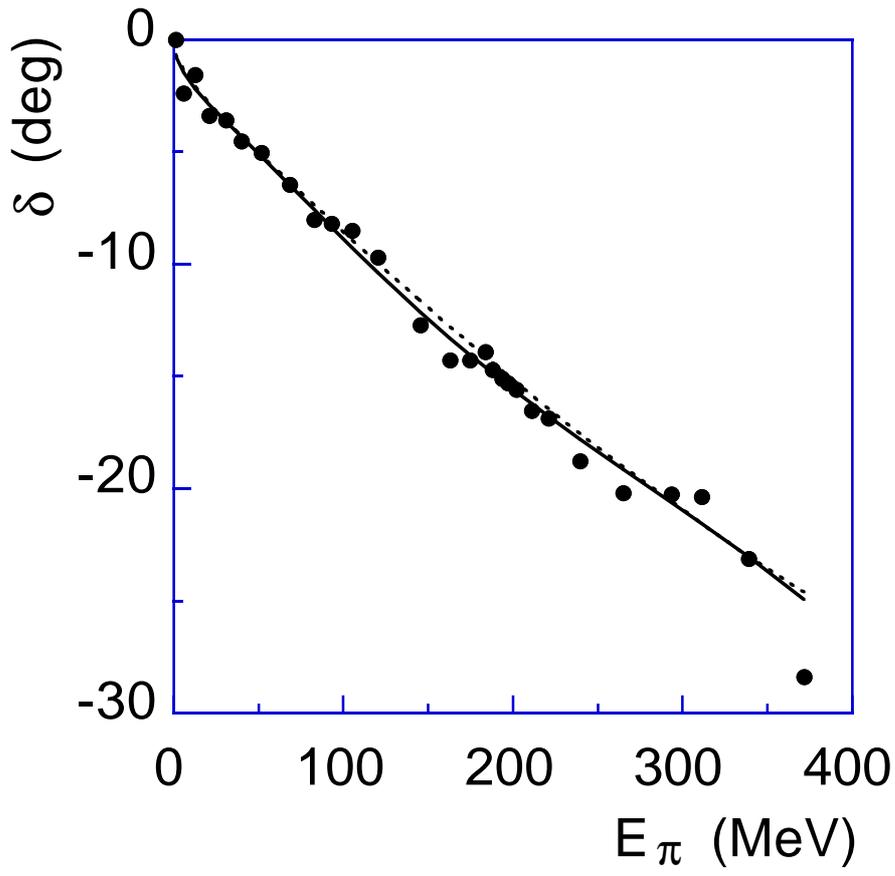
\begin{figure}
\caption{The resultant $\pi N$ phase shift. The solid line is obtained 
in this report and that used in Ref.~[1,2] for the dotted line. The 
empirical data are from Ref.~[5].}
\label{fig1}
\end{figure}

The parameters of this $S_{31}$ potential have been adjusted to fit the
scattering length and phase shift up to 400~MeV pion energy. The resultant
parameters are: $c = -0.0230136$, $\beta = 14.1251$, $M = 1589.375$ MeV, 
$\alpha_1 = 120.178$ MeV and $\alpha_2 = 240.863$ MeV.  
In Fig.~\ref{fig1} we compare the phase shifts for this potential (solid line)
with the phase shifts predicted by Thomas\cite{AWT76} and used in the previous
calculation\cite{SA94,SA95} (dotted line). Also included are the phase shift
based on the analysis of H\"ohler~{\it et al.}\cite{HK79} (dots). Clearly the
two sets of phase shifts are almost identical and consistent with the H\"ohler
{\it et al} analysis.  The scattering length for this new potential of
$a_3=-0.101\,m_\pi^{-1}$ is in better agreement with the empirical value of
$a_3=-0.101\pm0.004\,m_\pi^{-1}$\cite{KP80} when compared with the scattering
length of $a_3=-0.091\,m_\pi^{-1}$ for the potential of Ref.~\cite{AWT76}.

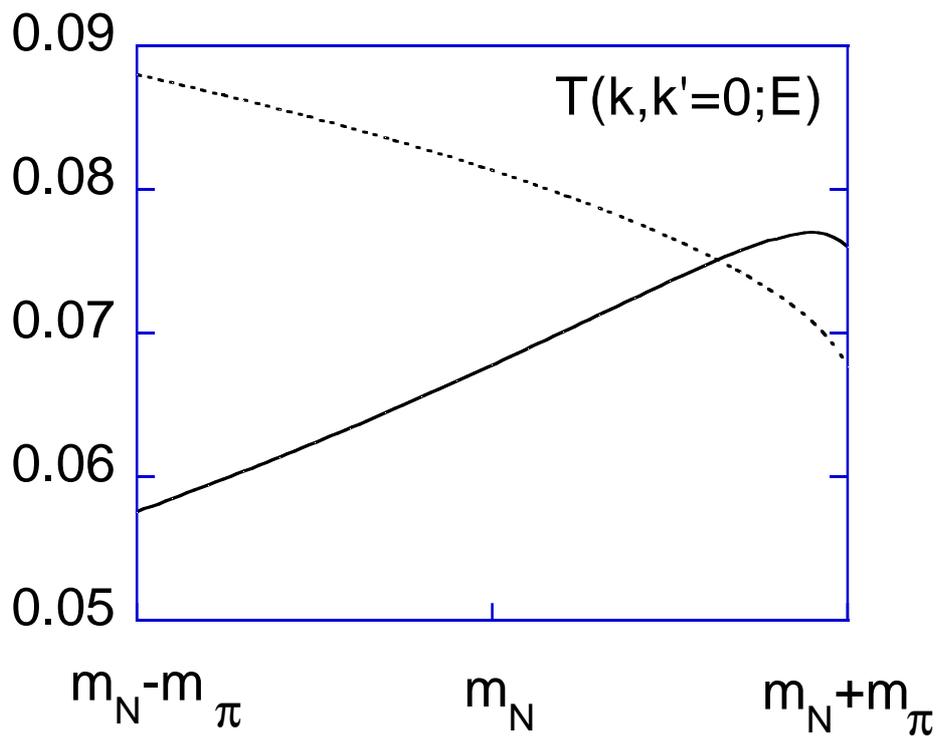
\begin{figure}
\caption{The energy dependence of the $\pi N$ t-matrix, $T(k,k';E)$, 
taking $k=k'=0$, for the $S_{31}$ channel. 
The solid line is obtained using the potential 
reported in this work, while the dotted curve is that used in Ref.~[1,2].}
\label{fig2}
\end{figure}

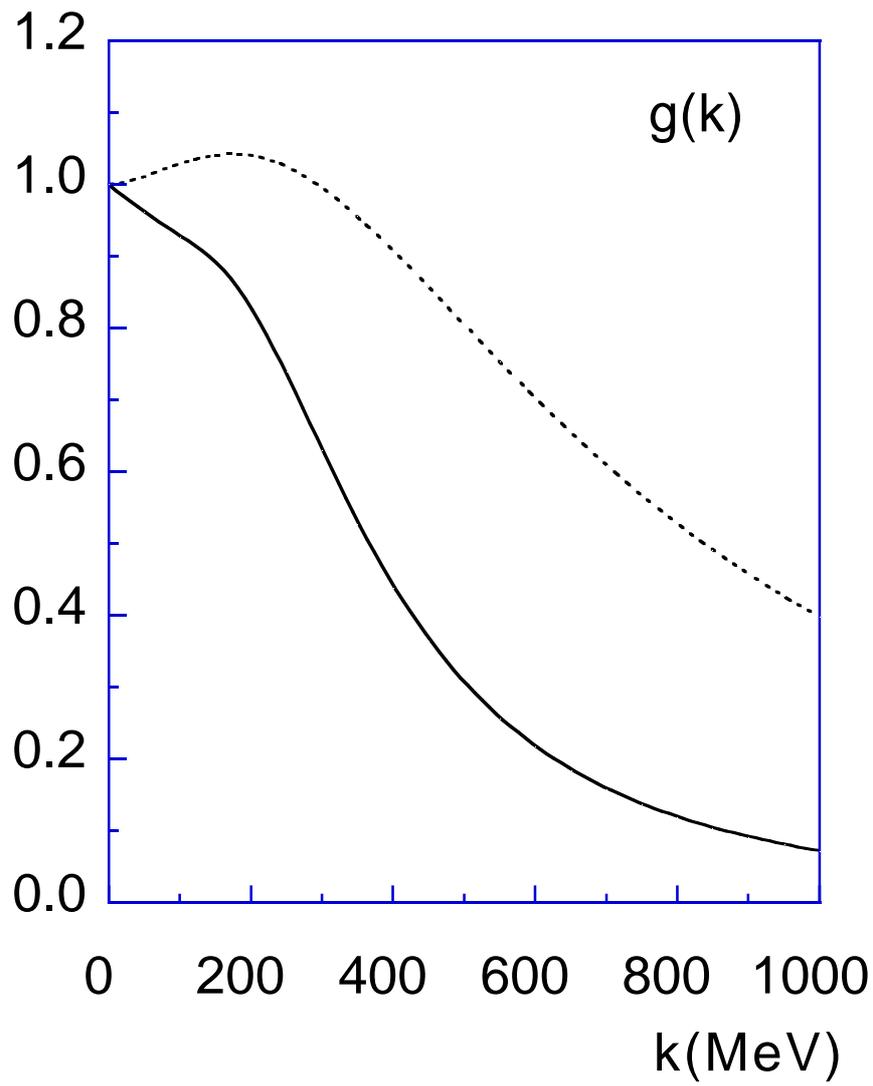
\begin{figure}
\caption{The solid line is the form factor $g(k)$, 
reported in this work, while the dotted curve is that used in Ref.~[1,2].}
\label{fig3}
\end{figure}

 In Fig.~\ref{fig2} we compare the subthreshold amplitude $T(k=0,k'=0;E)$ as
function of the energy $E$ for this potential (solid line), with that used in
Ref.~\cite{SA94,SA95}  (dotted line). Here we note that the t-matrix for this
new
potential has the positive slope and approaches zero as the energy decreases
while that of Ref.~\cite{SA94,SA95}  has a negative slope. As a result, 
we expect the contribution of the $S_{31}$  channel to the three-nucleon force
to
be reduced, and the total contribution from the $\pi-\pi$ three-body force
enhanced in comparison with the results reported previously\cite{SA94,SA95}. 
In Fig.~\ref{fig3} we compare the corresponding form factor for the $S_{31}$
potential (solid line) with that used in our previous calculation (dotted
line). This new form factor is considerably softer than that previously used.
However, it is important to note that this form factor is {\it not} the $\pi NN$
form factor associated with the pion production and absorption vertices. In
fact,
this form factor determines the off-shell behavior of the amplitude in
the $S_{31}$ channel, and to that extent is not constrained by the $\pi N$
scattering data, i.e. the phase shifts.

To examine the effect of changes to the subthreshold $\pi N$ amplitude in
the $S_{31}$ channel on the three-body force, we have calculated the
contribution of this partial wave to the three-nucleon force following the same
procedure as in Ref.~\cite{SA94,SA95}, {\it i.e.},  we use first order
perturbation theory. The triton wave function used  is obtained from the
solution
of the Faddeev equation with the  Paris-Ernst-Shakin-Thaler (PEST) potential
\cite{HP84,PK91}.  The $\pi NN$ form factor used is obtained from the $P_{11}$  
potential $PJ$ of Ref.~\cite{MA85}.  The resultant three-nucleon force
contribution is presented in  Table.~\ref{table1}. As we have expected, the
contribution of the $S_{31}$ partial wave is suppressed due to the smaller value
of the t-matrix and  the softer form factor. Here we note that the contribution
of the $S_{31}$ channel to the three-body force gets larger as we fix the
energy in the amplitude first at $E=m_N$ and then at $E=m_N+m_\pi$. This is due
to the  positive slope of $\tau(E)$ for this amplitude. Although this
reduction in the contribution of the $S_{31}$ amplitude has suppressed the
cancellation between the repulsive and attractive partial waves, and has given
a considerably more attractive contribution to the three-body force, the
overall contribution from the $P_{11}$ and $P_{33}$ channels of -8.8 and
-16.0~KeV is still too small to bridge the gap between the experimental
binding energy of 8.45~MeV and the commonly reported value for many of the
realistic potentials of 7.7~MeV\cite{FP93}. This could be due to the overall
energy dependence of the $\pi N$ amplitude and the range of the $\pi NN$ form
factor. The important result of this investigation is the fact that there is
considerable model dependence in the analytic continuation of the
$\pi N$ amplitude in the $S_{31}$ channel from the physical region to the
subthreshold region, and this introduces an element of uncertainty in the
determination of the three-nucleon force.


\acknowledgments

One of the authors (T-Y.S) is thankful for the hospitality 
of the Research Center for Nuclear Physics, Osaka University.

\begin{table}
\caption{The contribution of $S_{31}$ channel to the three-nucleon 
force are listed in the second and third columns. The resultant 
three-body force contribution (3BF) to the triton binding are 
listed in the fourth and fifth columns. We obtained those values 
by replacing only the $S_{31}$ contributions of Table 10 of Ref.[2].  
The second line is calculated taking the full energy dependence.
The third and fourth lines are  calculated by fixing the energy of 
$\tau(E)$ at $E=m_N$ and $E=m_N+m_\pi$, respectively. All energies are in keV.}
\label{table1}
\vskip 0.5 cm
\begin{tabular}{lcccc}
 & \multicolumn{2}{c}{$S_{31}$ contribution} & \multicolumn{2}{c}{3BF} \\
 & Ref.~[1,2] & This work & Ref.~[1,2] & This work \\  \tableline
 $\tau(E)$            & 31.7     & 4.1 & ~-2.3 & -29.9 \\
 $\tau(E=m_N)$        & 23.5     & 5.3 & -23.1 & -41.3 \\
 $\tau(E=m_N+m_\pi)$  & 19.5     & 5.9 & -55.5 & -69.1 \\
\end{tabular}
\end{table}

\end{document}